\documentclass[journal]{IEEEtran}

\usepackage{soul}
\usepackage{subfigure}
\usepackage{amsmath}
\usepackage{subfigure}
\usepackage{graphicx}
\usepackage{color}
\usepackage{dblfloatfix}
\usepackage[framemethod=tikz]{mdframed}

\newcommand{\fig}[1]{Fig.~\ref{fig:#1}}
\newcommand{\eq}[1]{(\ref{eq:#1})}

\ifCLASSINFOpdf
\else
\fi

\begin{document}
\markboth{Journal of \LaTeX\ Class Files,~Vol.~14, No.~8, August~2015}%
{Shell \MakeLowercase{\textit{et al.}}: Bare Demo of IEEEtran.cls for IEEE Journals}
\title{Flexible Raman Amplifier Optimization Based on Machine Learning-aided Physical Stimulated Raman Scattering Model}

\author{Metodi Plamenov Yankov,~\IEEEmembership{Member,~IEEE}, Francesco Da Ros,~\IEEEmembership{Senior Member,~OSA, Senior Member,~IEEE}, Uiara Celine de Moura,~\IEEEmembership{Member,~OSA}, Andrea Carena,~\IEEEmembership{Senior Member,~OSA, Senior Member,~IEEE}, and Darko Zibar%
\thanks{Metodi Plamenov Yankov, Francesco Da Ros, and Darko Zibar are with the Department of Electrical and Photonics Engineering, Technical University of Denmark, 2800 Kgs. Lyngby, Denmark. Uiara Celine de Moura was with the Department of Electrical and Photonics Engineering, Technical University of Denmark, 2800 Kgs. Lyngby, Denmark, and is now with NKT Photonics, 3460 Birker{\o}d, Denmark. Andrea Carena is with the Dipartimento di Elettronica e Telecomunicazioni, Politecnico di Torino, 10129 Torino - Italy. e-mail: meya@fotonik.dtu.dk}}

\maketitle

\begin{abstract}
The problem of Raman amplifier optimization is studied. A differentiable interpolation function is obtained for the Raman gain coefficient using machine learning (ML), which allows for the gradient descent optimization of forward-propagating Raman pumps. Both the frequency and power of an arbitrary number of pumps in a forward pumping configuration are then optimized for an arbitrary data channel load and span length. The forward propagation model is combined with an experimentally-trained ML model of a backward-pumping Raman amplifier to jointly optimize the frequency and power of the forward amplifier's pumps and the powers of the backward amplifier's pumps. The joint forward and backward amplifier optimization is demonstrated for an unrepeatered transmission of 250 km. A gain flatness of $<$ 1~dB over 4 THz is achieved. The optimized amplifiers are validated using a numerical simulator.  
\end{abstract}

\begin{IEEEkeywords}
Stimulated Raman scattering, Raman amplifier, frequency and power optimization, machine learning.
\end{IEEEkeywords}

\section{Introduction}
Raman amplifiers (RAs) present a significant advantage over erbium doped fiber amplifiers (EDFAs) in terms of noise figure and the potential for providing arbitrary gain profiles in a controlled way \cite{AgrawalRaman}. They are also a viable option for increasing the transmission bandwidth beyond the conventional C+L band and support S-band amplification \cite{SCL_Raman,Uiara2020multi}. Gain shaping over frequency is important for achieving uniformity of the quality of transmission for all channels in a wavelength division multiplexing (WDM) system. Gain nonuniformity is even more pronounced in ultra wide band systems due to different responses and losses of the optical components in different bands. Furthermore, in unrepeatered link scenarios, RAs are critical for providing amplification from the receiver end, as well as remote pumping \cite{CPQDUnrep2, CPQDUnrep,januario2019system}. Broadband amplification using RAs can be achieved by employing multiple Raman pumps at different frequencies. However, such configurations pose a challenge for the optimization of the pumps frequency and power due to the increased dimensionality of the problem \cite{RAopt}. This problem is exacerbated when higher order pumping is employed, as is often the case for unrepeatered systems. Genetic algorithms were proposed in \cite{RAopt, Neto2007}, which fall in the heuristic optimization methods category, and are prone to finding local optimums. State of the art approaches for optimization are also based on heuristics \cite{CPQDUnrep2,borraccini2022cognitive,della2019natural}. Machine learning (ML) methods for the optimization are also gaining traction \cite{RAmodel,soltani2022spectral,Chen2018,Zhou2006}, especially methods which train combinations of forward and inverse system models to predict the required pump power and frequency for a given target gain profile \cite{RAmodel, RAmodelUniPD, RAmodel1, mineto2021performance,Zibar2018}. ML methods provide excellent performance, however, they require a lot of training data to be generated in order to populate the 2$N_p$ dimensional space, where $N_p$ is the number of pumps. Furthermore, each training dataset is specific to $N_p$ and the number of wavelenght division multiplexed (WDM) channels to be amplified and does not allow re-optimization when pumps are to be added or when the data load changes, unless an expanded training dataset including all the configurations is considered~\cite{brusin2020loadaware}. 

This paper is an extension of our previous paper \cite{OECC}, where a completely flexible optimization method was proposed for forward-only pumping. In this paper, the forward propagation model from \cite{OECC} is combined with an ML-based model for the gain of a backward-pumping RA with a fixed number and frequency of the pumps in order to build a complete and differentiable model for a link with joint backward and forward pumping. The optimization is then completely flexible in terms of 1) number of pumps, their frequency and power for the forward RA; 2) power for the backward RA; and 3) fiber length (to a reasonable extent discussed later).  

The paper is summarized as follows. In Section~\ref{sec:model}, the physical SRS model and its forward propagation differentiable approximation are presented, including the ML-based model for the Raman gain coefficient. We also introduce the ML-based model for the gain of backward-pumping RA. In Section~\ref{sec:systems}, the unrepeatered systems under consideration are explained. In Section~\ref{sec:optimizations}, the optimization strategies are discussed. In Section~\ref{sec:results}, the optimized gain profiles are presented and analyzed. Section~\ref{sec:conclusion} concludes the work and provides an outlook.

\section{SRS modeling}
\label{sec:model}
The stimulated Raman scattering (SRS) effect in optical fibers results in transfer of power from carriers at high frequencies to carriers at low frequencies. The efficiency of this transfer depends on their respective carriers' frequency offset and their powers. It also depends on the type of fiber through its effective area. This effect allows for SRS-based amplification of communication signals. However, it also results in inter-carrier power transfer. At high powers, it results in undesired non-uniformity of the received power levels that will consequently define unbalanced level of optical signal to noise ratio (OSNR) for the different channels in the WDM system. The effect is exacerbated when wideband systems beyond the standard C-band are deployed for data transmission, e.g. C+L \cite{SRS}, or even S+C+L, as proposed in recent studies \cite{SCL_Raman}. 

The SRS can be described with a set of ordinary differential equations \cite{SRS}
\begin{align}
\label{eq:SRS}
 \frac{\partial P_n(z)}{\partial z} & = -2\alpha_n P_n(z) \notag \\ 
 & +\sum_{m=1}^{N} \frac{g_R(\omega_m-\omega_n)}{A_{eff}}{P_n(z) P_m(z)}
\end{align}
where $m, n \in \left[1; N\right]$ are the indexes of the carriers co-propagating through the fiber, $P_n(z)$ is the power at frequency index $n$ and distance $z$, $g_R$ is the Raman gain coefficient for a given offset between the angular frequencies $\omega_m$ and $\omega_n$, $\alpha_n$ is the fiber loss at the $n-$th frequency and $A_{eff}$ is the fiber effective area. In (\ref{eq:SRS}), no distinction is made between a 'pump' and a data-carrying 'channel'. Commonly, the efficiency of the power transfer, i.e.  $g_R(\omega)$, is characterized and tabulated for the more popular types of fiber, e.g. standard, single mode fiber (SSMF) \cite{Agrawal}.

Since there is no closed form analytical solution for Eq.~\eq{SRS}, finite difference numerical solutions are typically applied. Choosing a reference direction of propagation allows for the following expression to be formulated for the power evolution after a small propagation distance $\Delta_z$ \cite{SRS}
 \begin{align}
 \label{eq:apprSRS}
P_n(z) & = P_n(z-\Delta_z)-\alpha_n \Delta_z \notag \\
 & + \sum_{m=1}^{N} \frac{g_R(\omega_m-\omega_n)}{A_{eff}}L_{eff}(\Delta_z)e^{P_m(z-\Delta_z)},
\end{align}
where, $L_{eff}(L) = \frac{1-\exp(-2\cdot\alpha_n\cdot L)}{ 2\cdot \alpha_n}$ is the effective power interaction length. When all carriers propagate in the chosen reference direction, the solution \eq{apprSRS} converges to the true solution for \eq{SRS} for vanishing discretization step. However, the joint operation of forward and backward propagating carriers give rise to a non-trivial boundary value problem. In that case, iterative solvers must be employed for propagation in both directions of the fiber in order to arrive at an accurate solution, e.g. through a shooting algorithm \cite{liu2003shooting}. The forward-only solution is relevant for cases of inter-carrier SRS combined with a forward-only Raman pumping amplification. However, in the presence of backward pumping, the iterative solution needs to be applied. 

Optimization of any system requires that a cost function is formulated and minimized. In general, non-heuristic (e.g. gradient-based or analytical) optimization requires that the cost function is differentiable w.r.t. optimization parameters. In the case of RAs, that means an SRS model is needed in a form that is differentiable in the Raman pumps' power and frequency. First, the iterative nature of the solver for the general case of forward+backward propagating carriers results in non-differentiability in itself. 
Then, in \eq{apprSRS}, each step in $z$ is differentiable in power, but not in frequency due its dependence on $g_R$ which is typically described using a look-up table or piece-wise interpolation. These challenges are addressed in the following.

\subsection{Differentiable forward model}
\begin{figure}[!t]
\centering
 \includegraphics[trim=0 0 0 0cm, width=1.0\linewidth]{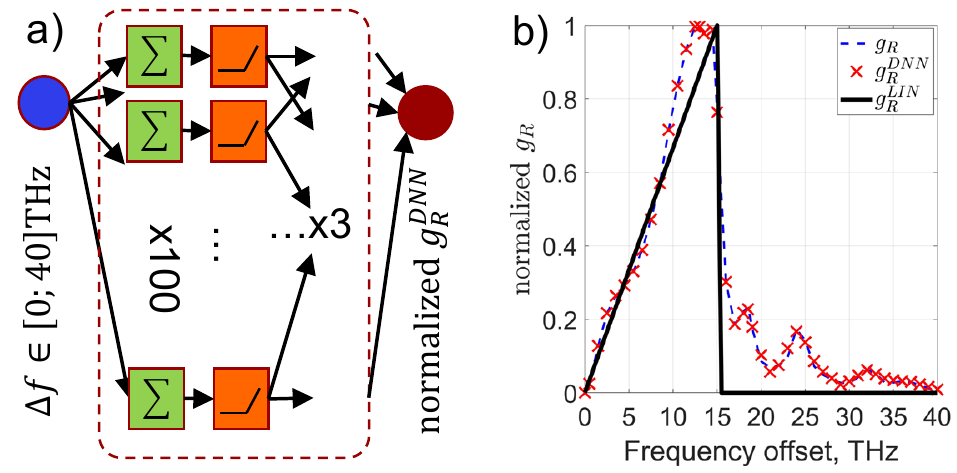}
 \caption{Neural network modeling of the normalized Raman gain coefficient $g_R(\omega_1-\omega_2)$. \textbf{a):} NN topology for modeling the coefficient: 3 layers, 100 nodes per layer, ReLU activation function; \textbf{b):} Modeling accuracy of the NN output $g_R^{DNN}$ compared to the true coefficient $g_R$ and the linear fit $g_R^{LIN}$ applied in \cite{UCL}.}
 \label{fig:gRNL}
%  \vspace{-0.3cm}
\end{figure}

In \cite{UCL}, $g_R$ is approximated using a linear interpolation $g_R^{LIN}$, which is differentiable. This is sufficient for estimating the SRS between WDM channels, but fails to provide the required accuracy when high-power pumps are added near the maximum efficiency (as will be demonstrated later in the paper). To that end, we train a deep neural network (DNN) to learn a nonlinear interpolation function $g_R^{DNN}$. The DNN is depicted in \fig{gRNL}a), it has 3 layers with 100 nodes per layer and a ReLU activation function and it is trained using gradient descent (GD) with the Adam optimizer and the mean squared error (MSE) cost function. In \fig{gRNL}b), the normalized $g_R^{DNN}$, $g_R^{LIN}$ and the true $g_R$ are given as a function of the frequency offset for standard, single mode fiber (SSMF). An MSE between $g_R$ and $g_R^{DNN}$ of $4.1\cdot 10^{-5}$ was achieved. The DNN is differentiable in the frequency offset and allows optimization w.r.t. pump frequency by substituting it for $g_R$ in \eq{apprSRS}. 

% The main drawback of this method is that it only allows optimization of forward Raman pumps, since solving (\ref{eq:SRS}) with backward-propagating pumps requires iterative algorithms which are not differentiable. 

\subsection{Differentiable backward model}
\label{sec:MLback}
The method for creating a differentiable model which includes backward pumping is detailed in \cite{RAmodel}. For a given fiber length $L_{span}$, number of backward pumps $N_{BCKW}$ and channel load $\{ P_1(0), P_2(0), \dots , P_N(0) \}$, ML can be applied to learn the on-off gain of the amplifier defined as
\begin{align}
 G(\omega_n)& = P_n(L_{span}, \{f_i^B, P_i^{B}, i=1\dots N_{BCKW}\}) \notag \\
            & - (P_n(0) - \alpha_n L_{span}),
\end{align}
where $P_n(L_{span}, \{f_i^B, P_i^{B}, i=1\dots N_{BCKW}\})$ is the output power of the fiber ($z=L_{span}$) at the desired frequency index $n$. The dependence on the $i-$th backward pump's frequency $f_i^B$ and power $P_i^B$ is also explicitly stressed. In order to train a model that predicts this mapping, a training data set is built based on either an iterative solver for the SRS, e.g. implementing a shooting algorithm, or experimentally. In either case, the training dataset is built by first sweeping the required free parameters, which can be any subset of $\{f_i^B, P_i^{B}\}, i=1\dots N_{BCKW}$. Then, recording $P_n(L_{span}, \{f_i^B, P_i^{B}\}, i=1\dots N_{BCKW})$ for all $n$. This method is very powerful, but suffers several important drawbacks. The main one is the dimensionality of the problem. Every time a pump is added to the dataset, due to the curse of dimensionality, the number of gain profiles that is needed for accurate representation of the parameter space grows exponentially. Furthermore, the model so constructed cannot generalize to predicting the gain profile for an arbitrary channel load when the amplifier is operated in saturation (high-input signal power leading to pump depletion)~\cite{brusin2020loadaware}. Nevertheless, it allows for an accurate representation of the gain profile of a backward pumping RA to be built, and provides sufficient accuracy in the un-saturated (pump undepleted) regime (see Section~\ref{subsec:FBA}). As in typical ML problems, the approximate gain profile $\{ G(\omega_n), n=1\dots N \}$ is predicted by an NN. The NN is fully differentiable, and so it allows gradient descent through it in order to optimize the pump configuration w.r.t. a desired gain profile. 

In this paper, we follow the process presented in \cite{RAmodel} where the NN model is trained using experimentally generated dataset. A 100 km of SSMF is used for transmission. The channel load is assumed fixed to 40 channels with 100 GHz spacing for a total input power of approx. 3~dBm, the pumps' frequencies are also fixed to $\left[206.1, 207.5, 209.0, 210.6\right]$ THz, respectively, and the pumps' powers are swept on a regular grid in the range (including boundary points) $\left[ 0; 21\right]$ dBm in order to record the on-off gain profile. Under these conditions the amplifier is operated outside the saturation regime and pump depletion is negligible. The NN then predicts the gain in dB for an arbitrary pump power configuration by essentially nonlinearly interpolating between the training gain profiles. The NN topology was coarsly optimized to 2 hidden layers of size 256 and 128, respectively with a ReLU activation function. The NN is trained to minimize the MSE 
\begin{align}
    MSE = {\mathrm{E}}_{n, i} \left[ |G_{meas}(\omega_n, i)-G_{model}(\omega_n, i)|^2 \right],
\end{align}
where $G_{meas}(\omega_n, i)$ and $G_{model}(\omega_n, i)$ are the measured and modeled gains at frequency index $n$ for pump profile $i$. The maximum absolute error (MAE), defined as
\begin{align}
MAE = \max_{n, i} \left[ |G_{meas}(\omega_n, i)-G_{model}(\omega_n, i)|\right], 
\end{align}
is also measured. The model achieved an MSE of 0.0329 dB$^2$ and a MAE of 1.0814 dB, similarly to~\cite{RAmodel}. For comparison,  a NN with the same topology was applied to a synthetic dataset generated using the GNPy software \cite{GNPy}. The achieved MSE and MAE for that dataset were 0.0033 dB$^2$ and 0.8892 dB, respectively. The slightly worse modeling performance of the experimental dataset is mainly attributed to the measurement uncertaintity in the experimental data.

%\begin{table}[!t]
% \caption{Summary of the NN parameters for the backward pumping amplifier.}
% \label{tbl:NNsum}
% \centering
% \begin{tabular}{c|c|c}
%  NN toplogy & MAE, dB & MSE, dB$^2$  \\
% \hline
%  [4x256x128x40] & 1.08 & 0.03 \\  
% \hline
% \hline
% \end{tabular}
% \vspace{-0.5cm}
%\end{table}

\section{Systems under consideration}
\label{sec:systems}
The RA optimization is tested on an unrepeatered link. Due to the long distance, such links require a very high transmit power. Here, forward RA can be applied to aid the transmitter erbium doped fiber amplifier (EDFA) and push some of the gain further into the fiber. This allows to keep a high signal power for longer compared to an EDFA-only setup. The RA then has a dual function: 1) to boost the power; 2) to shape the combined gain profile of the EDFA and the RA. Flattening the gain profile of the EDFA using standard methods, e.g. filtering is undesirable in such links due to the waste of power. In this work, without loss of generality, we assume an EDFA with a non-flat gain profile \cite{EDFAmodelPower}, resulting in a power profile launched into the fiber given in \fig{resultsFA_results} and \fig{resultsBA_results} for total launch powers of $15$~dBm and $18$~dBm, respectively. The total link distance we consider is 250 km. Similar to the backward model training stage, we consider a channel load of $N=40$ channels on a 100 GHz ITU grid. 

\subsection{Forward RA only with remote Raman pumping stage}
\begin{figure}[!t]
\centering
 \includegraphics[trim=0 0 0 0cm, width=1.0\linewidth]{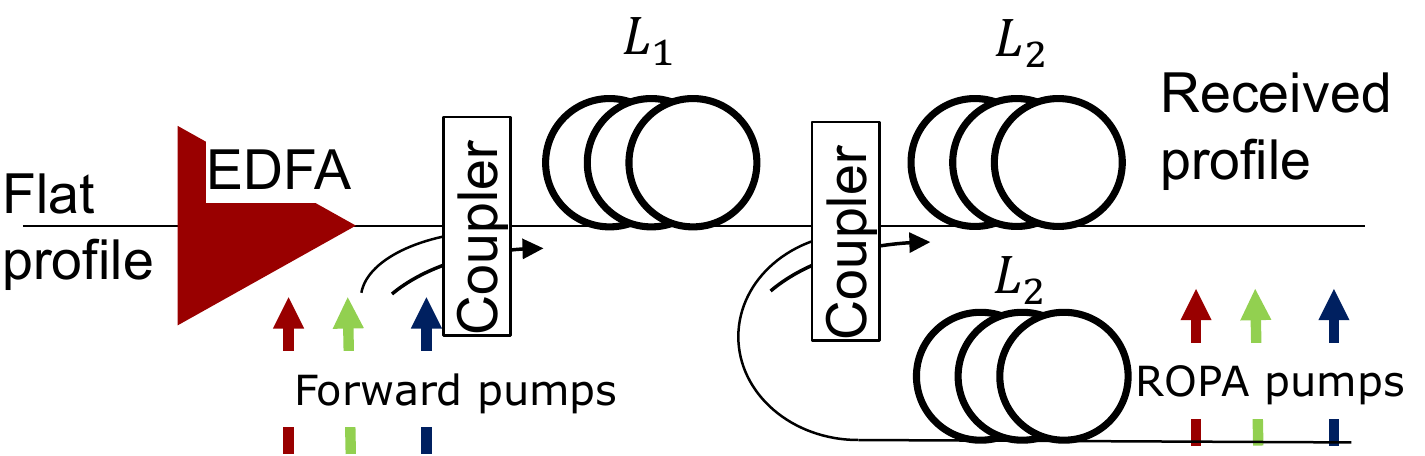}
 \caption{The unrepeatered transmission setup with a remote pumping stage. Both the transmitter and the remote RA are pumping in the forward direction.}
 \label{fig:setupFA}
%  \vspace{-0.3cm}
\end{figure}

The forward-only RA optimization is exemplified in \fig{setupFA}. In order to demonstrate the versatility of the model, we consider a Raman-based remote optically pumped amplifier (ROPA), where Raman pumps are guided from the receiver to the ROPA using an independent fiber, coupled into the transmission fiber and provide extra amplification for the last section of the span. In this case, both fibers $L_1$ and $L_2$ are modeled using \eq{apprSRS}, which is differentiable in the RAs frequency and power if $g_R^{LIN}$ or $g_R^{DNN}$ are applied. This system is motivated by \cite{bissessur201724}, where similar configuration is applied with forward RA and ROPA using a high-order Raman amplification. 

\subsection{Forward \& backward pumping}
\label{subsec:FBA}
\begin{figure}[!t]
\centering
 \includegraphics[trim=0 0 0 0cm, width=1.0\linewidth]{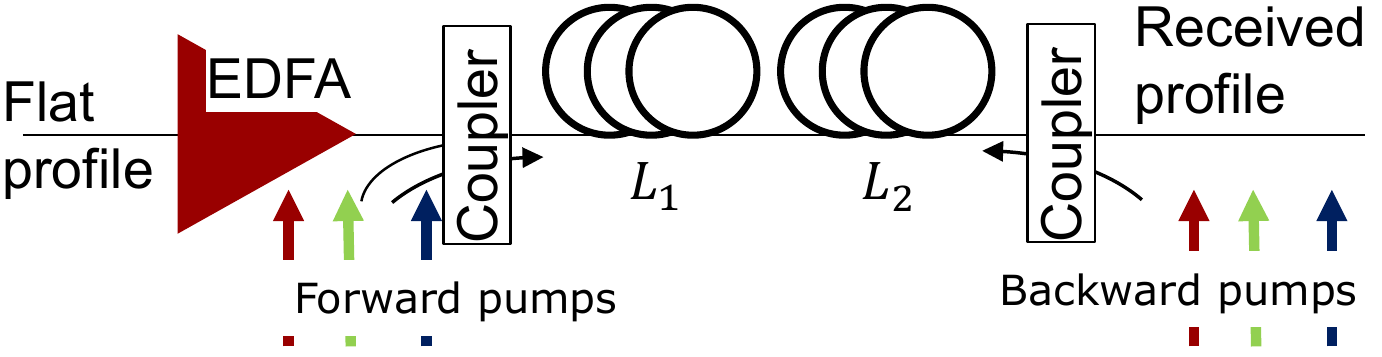}
 \caption{The unrepeatered transmission setup with both forward and backward pumping RA.}
 \label{fig:setupBA}
%  \vspace{-0.3cm}
\end{figure}

\begin{figure}[!t]
\centering
 \includegraphics[trim=0 0 0 0cm, width=1.0\linewidth]{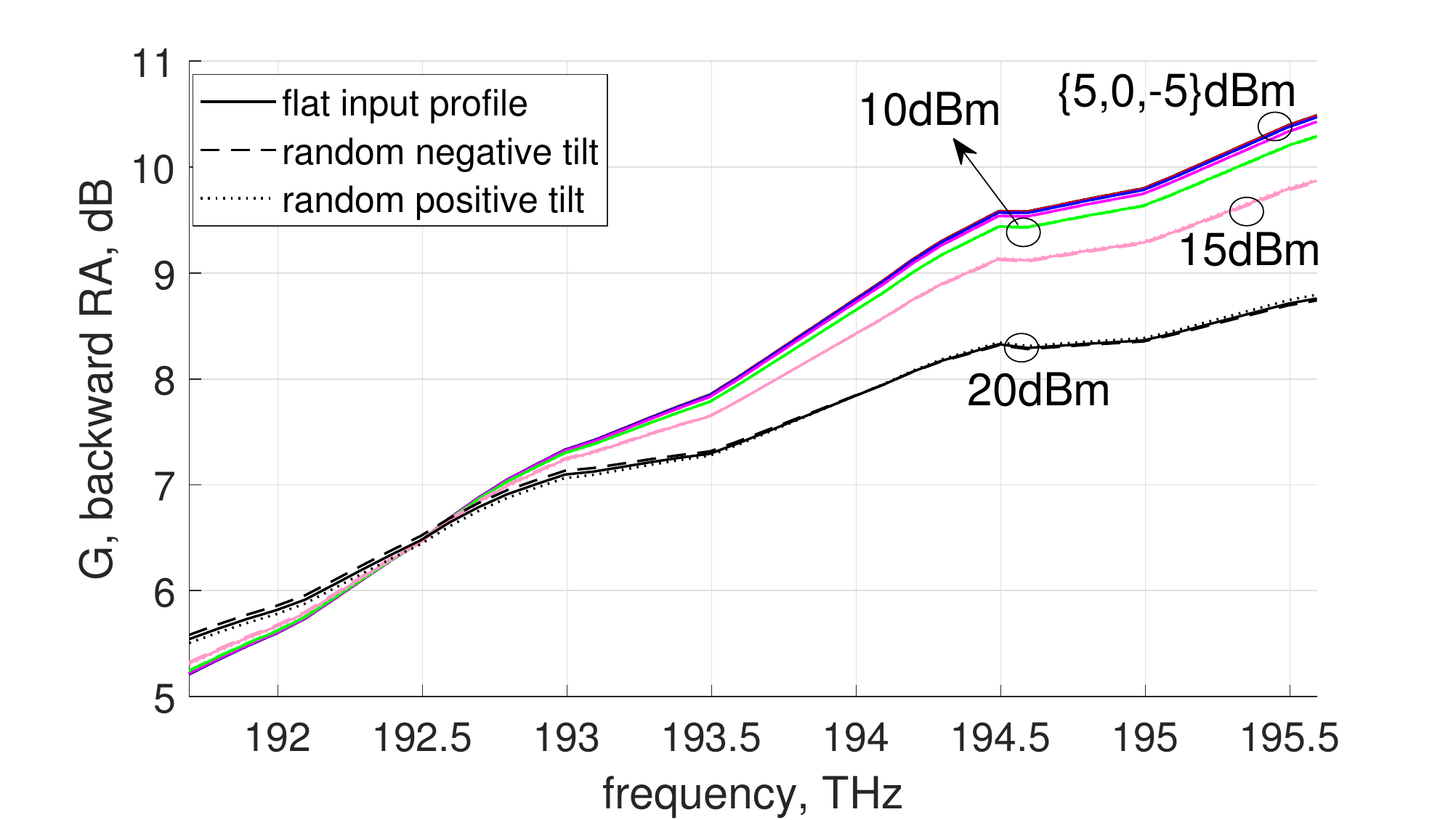}
 \caption{Depletion study for the backward RA. Slight dependence on the input channel load is seen for high total power ($20$ dBm). Below a total power of the load channels of $10$~dBm, the gain of the RA is independent of both the input power profile and the total power. Total power of the load profiles indicated with annotations.}
 \label{fig:depletionStudy}
%  \vspace{-0.3cm}
\end{figure}

The joint forward and backward RA optimization is exemplified in \fig{setupBA}. The receiver directly pumps in the backward direction. In this case, the distance is generally long enough and thus the signal power entering the backward-pumping section is weak enough to ensure that the pumps of the backward RA do not suffer from depletion. The un-saturated on-off gain then provides an accurate gain profile of the backward RA, and the ML model discussed in Section~\ref{sec:MLback} can be applied. In order to confirm this assumption, the difference in gain profiles for the backward RA is examined in \fig{depletionStudy}. The pump powers of all four backward pumps are fixed at their maximum of $21$~dBm. Different input load profiles at different total load powers (indicated with circles in the figure) are studied: a flat load profile, a randomly generated positively tilted load profile with an excursion of 5~dB and a randomly generated negatively tilted load profile with an excursion of 5~dB at total power input to a 100 km fiber of $\left[-5; 20\right]$~dBm. Below total powers of 10~dBm, the gain profile is constant and independent of both the input load profile and the total load power. This allows for the on-off gain model to be applied in unrepeatered systems. For such systems, after 100+ km of transmission, the total power will typically be well below the threshold of 10~dBm even if we apply a forward RA pumping in the first section of the span. The unrepeatered fiber span of length $L_1+L_2$ is modeled using two connected models of length $L_1$ and $L_2$. The first section is modeled using \eq{apprSRS}, and the second section is modeled using the ML model from Section~\ref{sec:MLback}. The concatenation of these models is completely differentiable in the forward RA's pump frequencies and powers, as well as in the backward RA's pump powers (pump frequencies are assumed fixed to the values presented in Section~\ref{sec:MLback}). 

\section{Optimization methods}
\label{sec:optimizations}
For the system with ROPA, the following cost function is adopted for the optimization of the RAs
\begin{align}
\label{eq:costForward}
 & L(\{P_i^F, f_i^F, P_j^R, f_j^R, i=1:N_{FRW}, j=1:N_{RMT}\}) = \\ \nonumber
 & \mathrm{E}_n \left[ |P_{modeled, n}(L_{span})-P_{target, n}(L_{span})|^2\right] + \\ \nonumber
 & ReLU(F_{min}-\min_j f_j^R) + ReLU(F_{min}-\min_i f_i^F) + \\ \nonumber
 & ReLU(\max_j f_j^R-F_{max}) + ReLU(\max_i f_i^F-F_{max}) + \\ \nonumber
 & ReLU(\max_j P_j^R-P_{max}) + ReLU(\max_i P_i^F-P_{max}) + \\ \nonumber
 & ReLU\left( \sum_j P_j^R-P_{tot} \right) + ReLU\left( \sum_i P_i^F-P_{tot} \right).
\end{align}
The first term in \eq{costForward} is the MSE over frequency between the modeled power $P_{modeled, n}(L_{span})$ and the target power $P_{target, n}(L_{span})$ , respectively. Then, $P_{max}$ is the power constraint per laser, $f_i^F$ and $f_i^R$ are the frequencies of the $i-$th forward and remote pump, respectively, $P_i^F$ and $P_i^R$ are the powers of the $i-$th forward and remote pump, respectively. The ReLU function are defined as $ReLU(x)=\max\left[0, x\right]$ and ensures the penalization of solutions outside of the power and frequency constraints of the pump lasers. Here, $F_{max}=220$ THz and $F_{min}=198$ THz are the frequency constraints with a range chosen to support 1st and 2nd order pumping for C-band operation.

For the system with backward pumping, the following version of the cost function is applied for optimization
\begin{align}
\label{eq:costBackward}
 & L(\{P_i^F, f_i^F, P_j^B, i=1:N_{FRW}, j=1:N_{BCKW}\}) = \\ \nonumber
 & \mathrm{E}_n \left[ |P_{modeled, n}(L_{span})-P_{target, n}(L_{span})|^2\right] + \\ \nonumber
 & ReLU(F_{min}-\min_i f_i^F) + ReLU(\max_i f_i^F-F_{max}) + \\ \nonumber
 & ReLU(\max_j P_j^B-P_{max}) + ReLU(\max_i P_i^F-P_{max}) + \\ \nonumber
 & ReLU\left( \sum_j P_j^B-P_{tot} \right) + ReLU\left( \sum_i P_i^F-P_{tot} \right).
\end{align}

After initial convergence, pump lasers with a frequency separation $\le 200$ GHz are merged together in a single pump with a power equal the sum of powers and frequency equal the average of the frequencies. The iterative optimization process is then resumed. This ensures a practical and efficient use of lasers limiting the number of pumps needed. Different variants can be devised of merging strategy, e.g. always selecting the $N$ pumps with highest powers and pruning the rest, or pruning based on minimum resulting deterioration of the MSE. In either case, the designer is free to impose a constraint on the number of pumps available for the RA and/or their frequency.

\section{Results}
\label{sec:results}
The optimization processes outlined in Section~\ref{sec:optimizations} are run for the systems under consideration with the target of a flat received power profile. In order to benchmark the optimizations, the optimized RAs are input to two reference implementations of a Raman solver.
\begin{enumerate}
    \item a full implementation in GNPy including a Raman solver for the set of ordinary differential equations describing the SRS (given in Eq.\eq{SRS}) and using a piece-wise interpolation for the tabulated $g_R$ values \cite{GNPy};
    \item a GNPy solver which assumes $g_R^{LIN}$ for the Raman efficiency. 
\end{enumerate}

\subsection{System with remote RA pumping stage}
\begin{figure}[!t]
 \centering
 \includegraphics[width=1.0\linewidth]{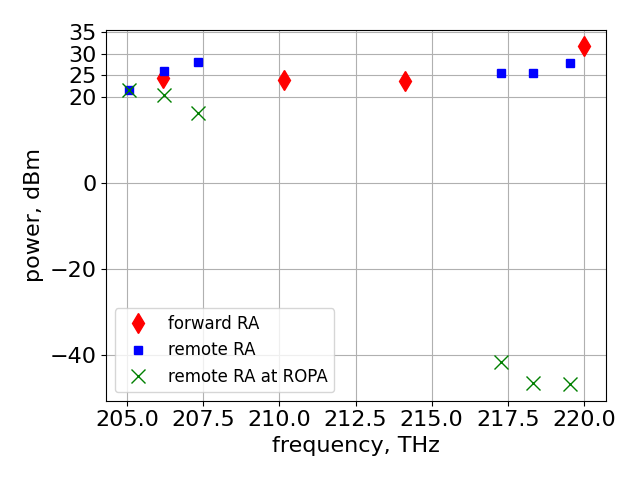}
\caption{Optimized Raman configuration when initialized to 20 remote and 20 forward pumps, converged to 6 remote and 4 forward pumps}
\label{fig:resultsFA_config}
\end{figure}

For the system given in \fig{setupFA}, the optimization algorithm was initialized with 20 pumps in each of the forward and the remote amplifier, $P_{max}= 2$~W and the EDFA is set to operate at an output power of $15$~dBm. The target total received power is chosen to be $-10$~dBm, or $\approx -26$~dBm per channel. In Fig.~\ref{fig:resultsFA_config}, the optimized forward and remote pumps are given. The model converged to $N_{FRW}=4$ and $N_{RMT}=6$, an MSE of $0.07 \text{dB}^2$ and an MAE of $0.67$ dB. The high-frequency pumps arrive at the ROPA at very low power, clearly indicating their role as second-order pumping for the low-frequency pumps. The latter then provide gain to the C-band. 

\begin{figure}[!t]
 \centering
 \includegraphics[width=1.0\linewidth]{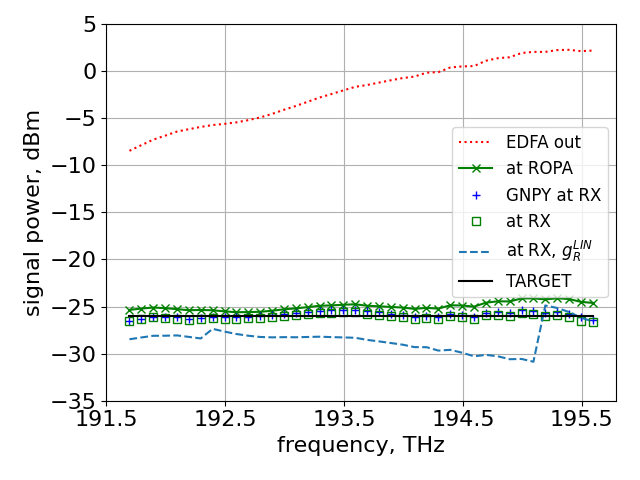}
\caption{Resulting power profiles at the beginning of the fiber ("EDFA out"), at the ROPA input and at the receiver ("at RX") for the RA configuration from \fig{resultsFA_config} (forward + ROPA), together with the GNPy benchmarks. }
\label{fig:resultsFA_results}
\end{figure}

The benchmarks are then given in Fig.~\ref{fig:resultsFA_results}. We see excellent correspondence between the model and the GNPy benchmark (maximum error of $0.7\text{dB}$ between the target profile and GNPy validation), which also justifies the chosen step size (see Section~\ref{sec:Discussion}). In this case, the forward RA provides $\approx 18.07$ dB of total gain, while the ROPA provides $\approx 6.88$ dB of total gain with total powers $\sum_i P_i^F=2.20$ W and $\sum_i P_i^R=1.52$ W, respectively. The amplifiers are optimized such that the forward RA flattens the power spectral density in the first section, while the ROPA provides a relatively flat gain. The linear assumption on $g_R$ results in significant discrepancies of $\approx 4.8$ dB of MAE, indicating that it is not feasible to apply this simple model for optimization. 

\subsection{System with backward pumping}
\begin{figure}[!t]
 \centering
 \includegraphics[width=1.0\linewidth]{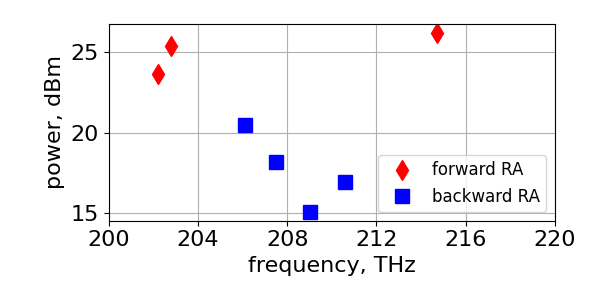}
\caption{Optimized Raman pumps' configuration for the system with backward pumping when initialized to 4 forward pumps, converged to 3 forward pumps}
\label{fig:resultsBA_config}
\end{figure}

\begin{figure}[!t]
 \centering
 \includegraphics[width=1.0\linewidth]{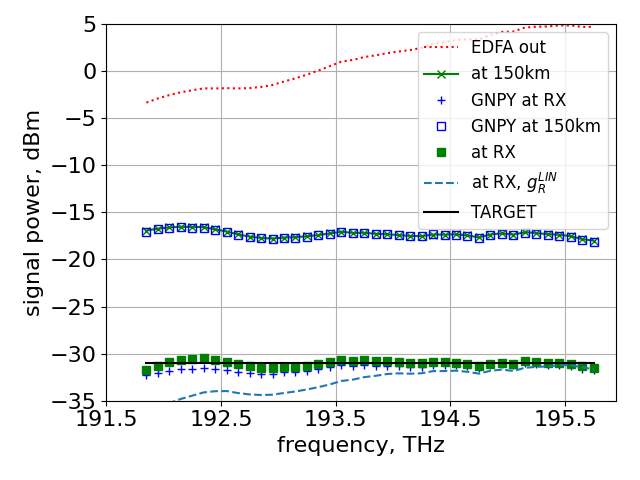}
\caption{Resulting power profiles at the beginning of the fiber ("EDFA out"), at the end of the first section ("at 150 km") and at the receiver ("at RX") for the RA configuration from \fig{resultsBA_config} (forward + backward), together with the GNPy benchmarks.}
\label{fig:resultsBA_results}
\end{figure}

For this system, the optimization algorithm was initialized with 4 pumps in the forward amplifier with $P_{max}= 1$~W. For the backward amplifier, $P_{max}= 0.14$~W, which is chosen to be within the operating region of the lasers used to generate the data for the ML model. The EDFA is set to operate at an output power of $18$~dBm. The target total received power is $-15$~dBm, or $\approx -31$~dBm per channel, which is realistic for such transmission scenarios. The numbers are chosen differently from the previous system to demonstrate the versatility of the optimization. In Fig.~\ref{fig:resultsBA_config}, the optimized forward and backward pumps are given. The model converged to $N_{FRW}=3$, an MSE of $0.08 \text{dB}^2$ and an MAE of $0.68$ dB. The benchmarks are then given in Fig.~\ref{fig:resultsBA_results}. A slightly higher maximum error of $1.25 \text{dB}$ between the target profile and GNPy validation of the proposed optimized profile is seen in this case. This is mainly attributed to potential mis-calibration between GNPY fiber parameters and the parameters of the fiber used to measure the on-off gain and train the ML model experimentally. This is confirmed by the fact, that the maximum error between the GNPy model (open squares) and the model from \eq{apprSRS} (green crosses) at the end of the first fiber section is $<0.1$~dB. The total power at the end of the first fiber section is $<0$~dBm, well below the threshold shown in \fig{depletionStudy} for depletion to occur. This assumption is further validated by the good correspondence to the GNPy benchmark at the receiver, which does not employ the two virtual sections but considers the fiber as a single span. In this case, the forward RA provides $\approx 9.25$ dB of gain, while the backward RA provides $\approx 5.25$ dB with total powers $\sum_i P_i^F=0.99$ W and $\sum_i P_i^B=0.21$ W, respectively. Similarly to the previous system, the resulting gain profile of the forward RA flattens the response of the EDFA, and the gain profile of the backward RA is relatively flat over the bandwidth. The linear assumption on $g_R$ results in significant discrepancies of $\approx 4.85$ dB of MAE.

\subsection{Final remarks}
\label{sec:Discussion}
The remaining discrepancies between the model from \eq{apprSRS} and the GNPy benchmark can potentially be further reduced by decreasing the step size at the expense of slower simulation time. At the chosen step size of $\Delta_z=100$m, the optimization requires $\approx 500$ iteration to converge which takes $\approx5$ minutes on a standard CPU. The discrepancies between the backward model and GNPy are as mentioned most likely due to remaining mis-calibration between the fiber parameters and non-modeled coupling losses, non-flat fiber loss over the C-band, the uncertainties in the measurements used to obtain the ML model, and similar. 

It should be noted that the forward-only pumping model is completely differentiable in distance $z$. This means that it supports shaping the RA gain over distance, which may be beneficial for systems, where power symmetry across the span is desired. Examples here include optical phase conjugation systems for nonlinearity compensation \cite{Minzioni}. 

\section{Conclusion}
\label{sec:conclusion}
Gradient-based optimization was proposed for the frequency and power of the pumps in Raman amplifiers. In the forward pumping case, a fully differentiable amplification model was proposed relying on machine learning (ML)-based nonlinear interpolation of the Raman gain coefficient. In the backward pumping case, a differentiable model is obtained using ML to train a NN representing the on-off gain of the amplifier. 

For the forward pumping case, this method is completely flexible in the main system and amplifier parameters and it allows for quick optimization of various system configurations, as exemplified with two different unrepeatered systems. The optimization follows the physical model of the SRS and does not require training data to be generated. An obvious extension is to include the RA and EDFA NFs in the model in order to optimize the received SNR profile instead of power similar to \cite{EDFAmodel}.

\section*{Acknowledgment}
The authors are supported by Danish National Research Foundation Centre of Excellence SPOC, DNRF123, the Villum YI, OPTIC-AI, 29344, the European Research Council (ERC-CoG FRECOM Grant No. 771878) and the Italian Ministry for University and Research (PRIN 2017, project FIRST). 

\bibliographystyle{IEEEtran}
\bibliography{references}

\end{document}